\documentclass[12pt]{article}

\def\lambar{{\mathchar'26\mkern-9.5mu\lambda}}

\title{\centerline{\normalsize physics/0201059 \hfill
SINP/TNP/02-3}\bigskip 
\bf Thermodynamics of a classical ideal gas\\ 
at arbitrary temperatures} 
\author{\bf Palash B. Pal\\ 
Theory Group, Saha Institute of Nuclear Physics\\
1/AF Bidhan-Nagar, Calcutta 700064}

\date{}

\begin{document}
\maketitle

\begin{abstract}\noindent
We propose a fundamental relation for a classical ideal gas that is
valid at all temperatures with remarkable accuracy.  All
thermodynamical properties of classical ideal gases can be deduced
from this relation at arbitrary temperature.
\end{abstract}

\bigskip

The famous equation of state for an ideal classical gas is
\begin{eqnarray}
PV = Nk_BT \,.
\label{PV}
\end{eqnarray}
Interestingly, a classical gas obeys this relation at all temperatures
as long as it is ideal, i.e., the Hamiltonian of the system does not
depend on the co-ordinates of the particles at all.  The proof is
simple, and appears in most textbooks \cite{texts}.  We will also
provide the proof later.

An equation of state, however, does not specify a system completely
\cite{callen}.  For example, from the equation of state in Eq.\
(\ref{PV}), we cannot find the entropy of the system, and many other
properties for that matter.  Of course if we have a fundamental
relation for the system, it contains all thermodynamic information
about the system including the equations of state themselves
\cite{callen}.  These are, for example, relations of the type
$S=S(U,V,N)$ or $U=U(S,V,N)$, which express the entropy or the
internal energy as a function of other extensive parameters of the
system.  Legendre transforms of these equations work just as well,
like the Helmholtz free energy $A$ as a function of $T$, $V$ and $N$.
However, it is often difficult to obtain such relations in closed
forms which would be valid for any temperature.  The Sackur-Tetrode
relation, for example, is a fundamental relation of the form
$S=S(U,V,N)$, but unlike the relation in Eq.\ (\ref{PV}), it is valid
only if the gas is non-relativistic, i.e., if the temperature is small
in the sense that $\beta mc^2\gg1$, where $m$ is the mass of the gas
particles.  Our aim in this article is to suggest a fundamental
relation for the classical ideal gas that can be used at any
temperature.

Since the gas is assumed to be ideal, the energy of any particle in
the gas depends only on its momentum.  At a momentum $p$, let us
denote the energy of a particle by $\varepsilon(p)$.  The
single-particle partition function in the canonical ensemble is then
given by
\begin{eqnarray}
Q = {gV \over (2\pi\hbar)^3} \int d^3p \; e^{-\beta \varepsilon(p)}
\equiv {g \over 2\pi^2 \lambar^3} \;  V f(\beta)\,,
%\label{}
\end{eqnarray}
where $g$ denotes the degeneracy due to internal degrees of freedom,
$\lambar$ is the Compton wavelength of the particle divided by $2\pi$,
\begin{eqnarray}
\lambar = {\hbar \over mc} \,,
%\label{}
\end{eqnarray}
and $f(\beta)$ is a dimensionless integral defined by
\begin{eqnarray}
f(\beta) = {1 \over (mc)^3} \int_0^\infty dp \; p^2 e^{-\beta
\varepsilon(p)} \,. 
\label{fdef}
\end{eqnarray}
The canonical partition function for a system of $N$ particles is then
given by
\begin{eqnarray}
Z = {Q^N \over N!} \,,
%\label{}
\end{eqnarray}
with the Gibbs correction factor for identical particles. This gives
\begin{eqnarray}
\ln Z = N \left[ \ln V + \ln f - \ln N + 1 + \ln (g/2\pi^2\lambar^3)
\right] \,, 
\label{lnZ}
\end{eqnarray}
where we have used Stirling's formula for $\ln N!$.  It is to be noted
that $f(\beta)$ is the only component which is not exactly known at
this point.  Our aim would be to determine it.

For this, we start by evaluating some thermodynamic quantities.  The
pressure of the gas is given by
\begin{eqnarray}
P = {1\over\beta} {\partial \ln Z \overwithdelims() \partial
V}_{\beta,N} = {N \over \beta V} \,,
\label{P}
\end{eqnarray}
irrespective of the functional form of $f(\beta)$.  This is Eq.\
(\ref{PV}), and this is why it is valid for any temperature.

Next we look at the energy density.  This is given by
\begin{eqnarray}
\varrho = - \frac 1V {\partial \ln Z \overwithdelims() \partial
\beta}_{V,N} = - {N \over V} {d \ln f\over d\beta} \,.
\label{rho}
\end{eqnarray}
Thus, if we define a quantity $w$ by
\begin{eqnarray}
P =  w\varrho \,,
\label{w}
\end{eqnarray}
Eqs.\ (\ref{P}) and (\ref{rho}) show that
\begin{eqnarray}
\frac1w = - \beta \, {d\ln f\over d\beta} = - {d\ln f\over d\ln \beta} \,.
\label{1/w}
\end{eqnarray}
Using the definition of $f(\beta)$ from Eq.\ (\ref{fdef}), we can then
write 
\begin{eqnarray}
w = {\displaystyle \int_0^\infty dp \; p^2 e^{-\beta \varepsilon(p)}
\over 
\displaystyle \int_0^\infty dp \; p^2 \beta \varepsilon(p) e^{-\beta
\varepsilon(p)}} \,.
\label{wp}
\end{eqnarray}
It is easy to see that if $\varepsilon(p)\propto p^\lambda$, this
formula gives $w=\lambda/3$, which gives the correct results for the
very low and very high temperature limits~\cite{texts}.

\begin{figure}
\centerline{\input{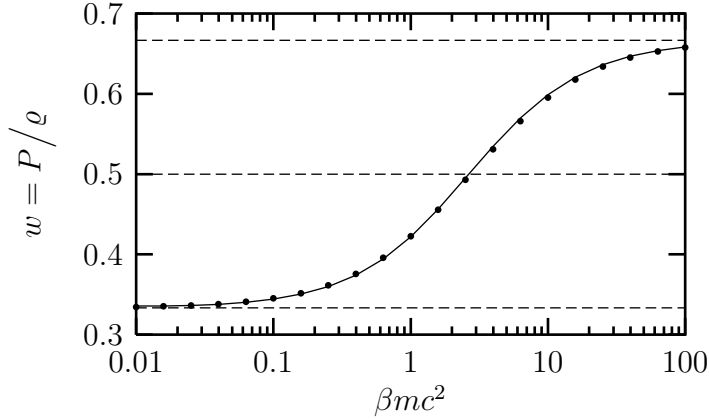}}
\caption{The ratio $P/\varrho$ for different temperatures.  The solid
line is obtained by numerical integrations from Eq.\ (\ref{wy}), and
the points correspond to the fit of Eq.\ (\ref{wfit}). The dashed
horizontal lines are drawn at $\frac13$, $\frac12$ and
$\frac23$.}\label{f:w} 
\end{figure}

For arbitrary temperatures, we should use the relativistically correct
formula for $\varepsilon(p)$.  For an ideal gas, there is no potential
energy.  The kinetic energy of a particle with momentum $p$ is given
by
\begin{eqnarray}
\varepsilon (p) = \sqrt{m^2c^4 + p^2c^2} - mc^2 \,.
%\label{}
\end{eqnarray}
Putting this into Eq.\ (\ref{wp}) and making a change of variable, we
can write 
\begin{eqnarray}
w = {\displaystyle \int_0^\infty dy
\; y^2 \exp \Big(- \xi \sqrt{1+y^2} \Big) 
\over \xi \displaystyle \int_0^\infty dy
\; y^2 \Big(\sqrt{1+y^2} - 1 \Big) \exp \Big(- \xi
\sqrt{1+y^2}\Big)} \,,
\label{wy}
\end{eqnarray}
where $\xi$ is a dimensionless variable for the inverse temperature:
\begin{eqnarray}
\xi = \beta mc^2 \,.
\label{defxi}
\end{eqnarray}
The integrations can be performed numerically for any value of $\xi$.
The results are shown by the solid line in Fig.~\ref{f:w}.
%%%%%%%%%%%%%%%%%%%%
%%%%%%%%%%%%%%%%%%%%

The shape of the curve is very similar to a hyperbolic tangent curve.
We can try a fit of the form
\begin{eqnarray}
w = \frac12 + \frac16 \tanh \Big(\alpha\ln(\xi/b) \Big) \,,
\label{tanh}
\end{eqnarray}
with two parameters $\alpha$ and $b$ which are both positive. 
This has the correct behavior for both extremes: $w=1/3$ for very
high temperatures ($\log\xi\to-\infty$) and $w=2/3$ for very low
temperatures ($\log\xi\to+\infty$). For reasons described below, we
must take $\alpha=\frac12$, which means that our fit is of the form
\begin{eqnarray}
w =\frac12 + \frac16 \; {\xi -b \over \xi +b} \,.
\label{wfit}
\end{eqnarray}
The choice for the parameter $b$ will be made later, in Eq.\
(\ref{b}).  However, in order to give a preview of the goodness of
this fit, we anticipate the value and show the resulting fit as points
in Fig.~\ref{f:w}.  As the figure suggests, the fit is very good for
all temperatures.  In fact, the agreement between the fit and direct
numerical calculation is considerably better than 1\% everywhere.

It has to be remarked that $\varepsilon(p)$, and consequently
$\varrho$, do not include the contribution from the rest mass energy
of the particles.  However, it is trivial to obtain a relation between
$P$ and the energy density $\rho$ which includes the mass energy as
well.  Obviously,
\begin{eqnarray}
\rho = \varrho + mc^2 n \,,
%\label{}
\end{eqnarray}
where $n=N/V$ is the number density of particles. Using Eq.\
(\ref{PV}), we can write it as
\begin{eqnarray}
\rho = \varrho + \xi P
\label{rho-varrho}
\end{eqnarray}
for a classical ideal gas.  If we then define an equation of state in
the form
\begin{eqnarray}
P = \omega \rho \,,
%\label{}
\end{eqnarray}
the constant $\omega$ will have a very simple relation with $w$ which
can be obtained by dividing both sides of Eq.\ (\ref{rho-varrho}) by
$P$: 
\begin{eqnarray}
{1\over \omega} = {1 \over w} + \xi \,.
%\label{}
\end{eqnarray}

We now go back to the fit of Eq.\ (\ref{wfit}), and notice that from
Eq.\ (\ref{1/w}), we can now determine $\ln f$.  In fact, using the
fit, we can write
\begin{eqnarray}
{d\ln f \over d \ln \xi} = -3 \; {\xi + b \over 2\xi + b} \,,
%\label{}
\end{eqnarray}
noting that $d\ln\beta=d\ln\xi$, since $mc^2$ is constant.  Thus,
\begin{eqnarray}
\ln f &=& -3 \int {d\xi \over \xi } \; {\xi + b \over 2\xi + b}
\nonumber\\ 
&=& K - 3 \ln \xi + \frac32 \ln (2\xi + b) \,,
\label{ffitk}
\end{eqnarray}
where $K$ is an integration constant.

To determine this constant, we may use of the form for $\ln f$ for
large $\xi$.  In this case, small momentum values dominate the
integrand, so that we can approximate $\varepsilon(p)$ by $p^2/2m$.
Once this is done, the integration in Eq.\ (\ref{fdef}) can be exactly
performed and one obtains
\begin{eqnarray}
f (\xi \gg1) \approx \sqrt{\pi \over 2} \; \xi^{-3/2} \,. 
\label{f<<1}
\end{eqnarray}
On the other hand, for $\xi\gg1$, Eq.\ (\ref{ffitk}) gives
\begin{eqnarray}
\ln f (\xi\gg1) \approx K - \frac32 \ln \xi + \frac32 \ln 2 \,.
%\label{}
\end{eqnarray}
Notice that this has the correct $\xi$-dependece as the
non-relativistic formula of Eq.\ (\ref{f<<1}).  This is the reason why
we had to take $\alpha=\frac12$ in Eq.\ (\ref{tanh}).  Moreover, this
also allows us to determine $K$, viz.,
\begin{eqnarray}
K = \frac12 \ln \pi - 2\ln 2 \,.
\label{K}
\end{eqnarray}
Putting this value of $K$ into Eq.\ (\ref{ffitk}), we obtain
\begin{eqnarray}
\ln f &=& \frac12 \ln \pi - 2\ln 2 - 3 \ln \xi + \frac32 \ln (2\xi +
b) \,.
\label{ffit}
\end{eqnarray}

\begin{figure}
\centerline{\input{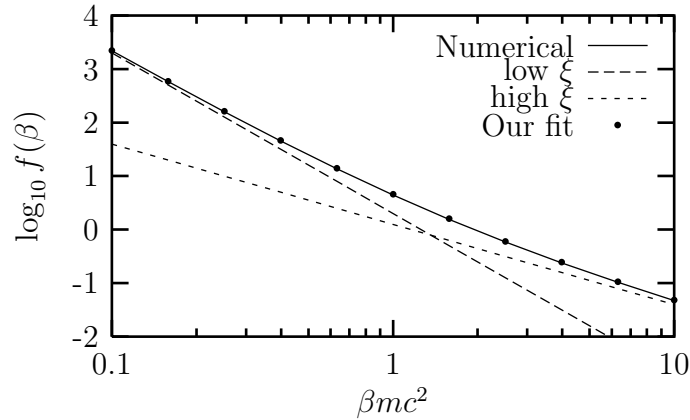}}
\caption{The integral $f(\beta)$ defined in Eq.\ (\ref{fdef}), plotted
as a function of $\xi=\beta mc^2$. The two dashed lines correspond to
the asymptotic forms for $f(\beta)$ given in Eqs.\ (\ref{f<<1}) and
(\ref{f>>1}).  The points are our fit from Eq.\ (\ref{ffit}), with the
value of $b$ taken from Eq.\ (\ref{b}).}\label{f:f} 
\end{figure}

The remaining constant $b$ can be determined by using the asymptotic
value of $f$ for small $\xi$.  In this case, large momentum values
dominate the integrand so that we can put $\varepsilon(p)=cp$.  Once
this approximation is made, the integration in Eq.\ (\ref{fdef}) can
be performed analytically and we obtain
\begin{eqnarray}
f(\xi\ll1) \approx 2 \xi^{-3} \,.
\label{f>>1}
\end{eqnarray}
On the other hand, Eq.\ (\ref{ffit}) gives
\begin{eqnarray}
\ln f(\xi\ll1) \approx 
\frac12 \ln \pi - 2\ln 2 - 3 \ln \xi + \frac32 \ln b \,.
%\label{}
\end{eqnarray}
Comparing the two expressions, we obtain
\begin{eqnarray}
b = {4 \over \sqrt[3]\pi} \,.
\label{b}
\end{eqnarray}
This finally completes the fit that we sought for. The fit is shown in
Fig.~\ref{f:f}, along with the values obtained by direct numerical
integration.

%%%%%%%%%%%%%%%%%%%%
%%%%%%%%%%%%%%%%%%%%
We would like to make a comment about our choice of the parameter
$b$.  We have chosen it so as to fit the low-$\xi$ end perfectly.
However, the resulting fit is worst (though still better than 1\%) for
intermediate values of $\xi$, which shows clearly in Fig.~\ref{f:w}.
If we take a slightly smaller value of $\xi$, the middle part fits
much better.  Of course it also means that the agreement for small
$\xi$ gets a bit worse.  To be precise, if instead of the value $b$
given in Eq.\ (\ref{b}) we choose some other value $b'$ for the
parameter under consideration, the values of $\ln f$ mismatches by an
amount $\frac32 \ln(b/b')$ for $\xi\ll1$.  But since the value of
$f(\beta)$ itself is very large for small values of $\xi$, the
fractional error is very small.  A choice like $b=e$, the base of
natural logarithms, makes the overall fit much better, to better than
$0.5\%$, for the entire range of temperatures.  If we go further down,
say $b=8/3$,  the fit becomes very good in the middle but considerably
worse at the ends.

We can put our fit for $f(\xi)$ into Eq.\ (\ref{lnZ}).  This gives
\begin{eqnarray}
\ln Z = N \left[ \ln {gV  \overwithdelims() N \lambar^3} 
+ \frac32 \ln {2\xi + b  \overwithdelims()
4\pi \xi^2} + 1 \right] \,,
\label{lnZfit}
\end{eqnarray}
where $b$ is given in Eq.\ (\ref{b}).  The Helmholtz free energy of
the classical ideal gas is then given by
\begin{eqnarray}
A = - {N \over \beta} \left[ \ln {gV  \overwithdelims() N \lambar^3} 
+ \frac32 \ln {2\xi + b  \overwithdelims()
4\pi \xi^2} + 1 \right] \,.
\label{Afit}
\end{eqnarray}
If we recall the definition of $\xi$ from Eq.\ (\ref{defxi}), we
realize that this is an expression for the free energy as a function
of the volume, temperature and total number of particles.  This is
thus a fundamental relation which can be used at any temperature.  All
thermodynamic properties of an ideal gas can then be deduced from this
relation at arbitrary temperature.

This is the result.  Of course it is not exact; it is a fit to the
numerical results.  But it is a very good fit, and the usefulness of
such an expression cannot be overemphasized.  In most practical
situations, a closed analytic expression is much easier to use than
numerical integrations.  Our expression for the Helmholtz free energy
in Eq.\ (\ref{Afit}) provides such an analytic expression which can be
used at any temperature to obtain all thermodynamic properties of a
classical ideal gas with suitable manipulations.

We can of course write the fundamental relation in alternative but
equivalent ways.  For example, we can write the fundamental relation
for the entropy $S$ by expressing it as a function of $U$, $V$ and
$N$.  For this, we use Eq.\ (\ref{Afit}) to obtain
\begin{eqnarray}
{S \over Nk_B} = - {1 \over Nk_B} 
{\partial A \overwithdelims() \partial T}_{V,N} 
= \ln {gV  \overwithdelims() N \lambar^3} 
+ \frac32 \ln {2\xi + b  \overwithdelims()
4\pi \xi^2} + {5\xi+4b \over 2\xi+b} \,.
\label{S:VNxi}
\end{eqnarray}
At this point, the right hand side is a function of $V$, $N$ and $T$
(through $\xi$). To put it in the form of a fundamental relation, we
must eliminate $\xi$ and bring in the internal energy $U$.  This can
be done through the fit of Eq.\ (\ref{lnZfit}), which gives
\begin{eqnarray}
U = - {\partial \ln Z \overwithdelims() \partial \beta}_{V,N} = 3Nmc^2
\; {\xi+b \over \xi(2\xi+b)} \,.
\label{U:Nxi}
\end{eqnarray}
Recalling the definition of $\xi$ from Eq.\ (\ref{defxi}), it can be
easily checked that this gives $U=3N/2\beta$ in the non-relativistic
limit ($\xi\gg1$), and $U=3N/\beta$ in the ultra-relativistic limit
($\xi\ll1$).  In the general case, we can solve Eq.\ (\ref{U:Nxi}) to
obtain 
\begin{eqnarray}
\xi = {3Nmc^2 -bU + \sqrt{(3Nmc^2 -bU)^2 + 24bUNmc^2} \over 4U} \,.
\label{xi:U}
\end{eqnarray}
Putting this back into Eq.\ (\ref{S:VNxi}), we obtain the expression
for entropy as a function of the external variables $U$, $V$ and $N$.
As stated before, this is also a fundamental relation.  The
non-relativistic limit should now be acknowledged as $U\ll Nmc^2$,
where Eq.\ (\ref{xi:U}) gives $\xi=3Nmc^2/2U$, which is large.
Putting this solution into Eq.\ (\ref{S:VNxi}), we obtain
\begin{eqnarray}
{S \over Nk_B} &=& \ln {gV  \overwithdelims() N \lambar^3} 
+ \frac32 \ln {U\overwithdelims()
3\pi Nmc^2} + {5 \over 2} \nonumber\\
&=& \ln \left[ {m \overwithdelims() 3\pi\hbar^2}^{3/2} \; {gVU^{3/2}
\over N^{5/2}} \right] + {5 \over 2} \,,
%\label{}
\end{eqnarray}
which is the Sackur-Tetrode relation \cite{texts}.  In the other
extreme, when $U\gg Nmc^2$ and so $\xi=3Nmc^2/U\ll1$, we obtain
\begin{eqnarray}
{S \over Nk_B} 
&=& \ln \left[ {\pi \over (3\pi\hbar c)^3} \; {gVU^3
\over N^4} \right] + 4 \,,
%\label{}
\end{eqnarray}
if we use the value of $b$ from Eq.\ (\ref{b}).  This result can also
be obtained analytically by using $\varepsilon=cp$.

I thank R.~K. Moitra for the suggestion of including the expression
for entropy in the paper.

%%%%%%%%%%%%%%%%%%%%


\begin{thebibliography}{[W]}
\bibitem{texts} Some of my favorite textbooks are:
%
\begin{list}{\alph{enumii})}{\usecounter{enumii}}
\item F. Reif: {\sl Fundamentals of statistical and thermal physics},
McGraw-Hill, 1965;
\item E. M. Lifshitz and L. P. Pitaevskii: {\sl Statistical physics},
Part 1, Pergamon Press, 3rd edition, 1980;
\item R. K. Pathria: {\sl Statistical mechanics}, Pergamon Press, 1972.
\end{list}
%

\bibitem{callen} H. B. Callen, {\sl Thermodynamics}, John Wiley and
sons, 1960.  Most other thermodynamics textbooks do not emphasize
these issues clearly.

\end{thebibliography}
\end{document}